\documentclass[preprint,5p,times,twocolumn]{elsarticle}

\usepackage{amsthm, amsfonts, amssymb}	
\usepackage{epstopdf}        			
\usepackage[fleqn]{amsmath}				
\usepackage{graphicx}					
\usepackage{txfonts}					
\begin{document}

\begin{frontmatter}
\title{
Bohmian mechanics and Fisher information
	   for $q$-deformed Schr\"odinger equation
}

\author[sertao]{Bruno G. da Costa\corref{cor1}}
\ead{bruno.costa@ifsertao-pe.edu.br}
\author[capes]{Ignacio S. Gomez}
\ead{nachosky@fisica.unlp.ar}

\cortext[cor1]{Corresponding author}
\address[sertao]{Instituto Federal de Educa\c{c}\~ao, Ci\^encia e Tecnologia do Sert\~ao Pernambucano,
				 BR 407, km 08, 56314-520 Petrolina, Pernambuco, Brazil}

\address[capes]{Instituto de F\'{i}sica, Universidade Federal da Bahia,
			    Rua Barao de Jeremoabo, 40170-115 Salvador-BA, Brazil\\
				National Institute of Science and Technology for Complex Systems, Brazil\\
				IFLP, UNLP, CONICET, Facultad de Ciencias Exactas,
				Calle 115 y 49, 1900 La Plata, Argentina}

\begin{abstract}
We discuss the Bohmian mechanics by means of the deformed
Schr\"odinger equation for position dependent
mass, in the context of a $q$-algebra inspired by nonextensive statistics.
A deduction of the Bohmian quantum formalism is performed
by means of a deformed Fisher information functional, from which a deformed
Cram\'er-Rao bound is derived.
Lagrangian and Hamiltonian formulations, inherited by the
$q$-algebra, are also developed.
Then, we illustrate the results with a particle confined
in an infinite square potential well.
The preservation of the deformed Cram\'er-Rao bound for
the stationary states shows the role played by the $q$-algebraic structure.
\end{abstract}


\end{frontmatter}


\section{\label{sec:intro}Introduction}

Along several decades it has been shown that fundamental disciplines
can be treated as theories of inference, where the available information about
the system allows one to derive the dynamics from making use of probability theory.
Between the most important methods of inference the maximum entropy one
is found, where a rule is given (typically, the maximization of a functional $S[\rho]$)
for obtaining the distribution $\rho$
that represents the best knowledge of the system constrained by the available
information \cite{jaynes_1957}.
When $S[\rho]$ is chosen to be the Shannon-Gibbs entropy then the so-called
MaxEnt method results.
In particular, a functional of interest is the Fisher information (FI) $I_F[\rho]$,
which measures the information of an observable variable of $\rho$ with respect
an unknown parameter associated.
The FI satisfies an important rule called the Cram\'er-Rao bound,
by stating a lower bound for the covariance determinant of the variables
in terms of the estimated parameters.
The FI can be used to derive the quantum and relativistic mechanics
by means of variational principles,
where the constraints contain the physics \cite{Frieden-book}.

In this context, an interesting application of MaxEnt
and FI is the deduction of the Bohmian quantum formalism
\cite{Reginatto-1998,Hall-Reginatto-2002}, which was
introduced by Bohm \cite{Bohm-1952} as an alternative interpretation of
the quantum mechanics using the idea of the {\it de Broglie pilot wave}
\cite{deBroglie-1927}.
Plastino {\it et al.} at
\cite{Plastino-Casas-Plastino-2001} studied Hamiltonians
with a position dependent effective mass,
which are widely used in many areas,
both experimentally and theoretically:
semiconductors \cite{vonroos_1983},
quantum dots \cite{Serra-Lipparini-1997},
many body theory \cite{Bencheikh-et-al-2004},
superintegrable systems \cite{Ranada-2016},
quantum liquids \cite{Barranco-1997},
inversion potential for $\textrm{NH}_3$ \cite{Aquino_1998},
astrophysics \cite{Richstone_1982},
nonlinear optics \cite{Li-Guo-Jiang-Hu},
relativistic quantum mechanics \cite{Alhaidari-2004},
nuclear physics \cite{Alimohammadi-Hassanabadi-Zare-2017},
etc.

In the mathematical description of quantum systems
with position-dependent effective mass, the mass operator $m(\hat{x})$
and the linear momentum $\hat{p}$ are not commutating.
A general form for the Hermitian kinetic energy operator
has been suggested by O. von Roos \cite{vonroos_1983}
which characterizes the most of those used in the literature.
The ordering problem of the kinetic energy operator has been investigated by:
BenDaniel and Duke \cite{BenDaniel-Duke-1966}, Gora and Williams \cite{Gora-Williams-1969},
Zhu and Kroemer \cite{Zhu-Kroemer-1983}, Li and Kuhn \cite{Li-Kuhn-1993}.
Recently, a $q$-deformed Schr\"odinger equation
associated with a position-dependent mass
\cite{costa-filho-2011,costa-filho-2013,mazharimousavi,costa-borges}
has been studied in the context of a generalized translation
operator related to a nonextensive algebraic structure \cite{borges_2004}.

Basing us in previous works \cite{costa-filho-2011,costa-filho-2013,mazharimousavi,costa-borges}
and applying the variational principle to a $q$-deformed version of
the FI inspired in nonextensive statistics,
in this paper we discuss a $q$-deformed Bohmian quantum theory
associated with the $q$-deformed Schr\"odinger equation, along with
the corresponding Lagrangian and Hamiltonian formulations.
Also, we derive a Cram\'er-Rao bound associated with the FI proposed.

The paper is organized as follows. In Section \ref{sec:q-SE} we review
the deformed Schr\"odinger equation for position dependent effective mass.
Section \ref{sec:q-Bohm-theory} is devoted to a deformed Bohmian quantum theory based
on the deformed Schr\"odinger equation and using the
de Broglie wave-pilot interpretation. Next, in Section \ref{sec:q-fisher}
we present a deformed Fisher functional for a position-dependent mass system.
Here we deduce a Cram\'er-Rao bound associated with the deformed FI.
Then, in Section \ref{sec:appl} we illustrate the results with
a particle confined in an infinite square potential well.
For comparing, we calculate the deformed
Cram\'er-Rao bound and the standard one for some stationary states.
Finally, in Section \ref{sec:conclusion} we draw some conclusions
and future directions are outlined.

\section{\label{sec:q-SE}
		Review of the $q$-deformed Schr\"odinger equation
	    for position-dependent mass}

Nonextensive statistical mechanics constitutes a formalism
of wide applicability in several areas of physics
\cite{Entropia-Tsallis,Tsallis-book-2009}.
The mathematical background of this approach is based on the generalized functions:
the $q$-exponential,
$
	\exp_q (u) \equiv [1 + (1-q)u]_{+}^{1/(1-q)},
$
with $[A]_{+} = \textrm{max} \{ A, 0\}$,
where the parameter $q$ (the so-called entropic index) has been found to have several
physical interpretations \cite{Tsallis-book-2009}.
Considering the $q$-addition and $q$-subtraction operators
defined respectively by $a \oplus_q b \equiv a + b + (1-q)ab$
and  $a \ominus_q b \equiv \frac{a-b}{1 + (1-q)b}$
($b \neq \frac{1}{q-1}$), the $q$-exponential
satisfies $\exp_q(a) \exp_q(b) = \exp_q(a\oplus_q b)$
and $\exp_q(a)/\exp_q(b) = \exp_q(a\ominus_q b)$.
Moreover, as proposed in \cite{borges_2004},
from the definition of the $q$-deformed infinitesimal element
\begin{equation}
	d_q u = \lim_{u'\rightarrow u} u' \ominus_q u = \frac{du}{1+(1-q)u},
\end{equation}
one can define the $q$-deformed derivative operator
\begin{equation}
	D_q f(u) = \lim_{u'\rightarrow u} \frac{f(u')-f(u)}{u'\ominus_q u}
			 = [1 + (1-q)u]\frac{df}{du},
\end{equation}
and the $q$-deformed integral
\begin{equation}
	\int f(u) d_q u = \int \frac{f(u)}{1 + (1-q)u}du.
\end{equation}
These operators satisfy the properties
$D_q \exp_q (u) = \exp_q (u)$ and
$\int \exp_q (u) d_q u = \exp_q (u) + \textrm{constant}$.

Recently, Costa Filho {\it et al.}
\cite{costa-filho-2011,costa-filho-2013,mazharimousavi,costa-borges}
have introduced a generalized translation operator
which produces nonadditive
spatial displacements, i.e.,
\begin{equation}
\label{eq:T_gamma}
	\hat{\mathcal{T}}_{\gamma}(\varepsilon)|x \rangle
			= | x + \varepsilon + \gamma x \varepsilon \rangle
\end{equation}
where $\varepsilon$ is an infinitesimal displacement and $\gamma$
is a parameter with dimension of inverse length
whose physical role is as follows.
If $L_0$ is the characteristic volume of the system,
then by defining $\gamma_q \equiv (1-q)/\xi,$
where $\xi$ is a characteristic length such that
$\gamma_q L_0 \sim 1$
(i.e., $1 - q \propto \xi/L_0$) then
$1-q$ can be interpreted as a coupling measurement between $\xi$ and $L_0$.
The lower the ratio $\xi/L_0$, the closer the parameter $q$ should be to 1.
Thus, the right hand side of (\ref{eq:T_gamma}) can be identified as
the $q$-addition $\xi[({x}/{\xi}) \oplus_q ({\varepsilon}/{\xi})]$.
The usual case is recovered for $q \rightarrow 1$ ($\gamma_q \rightarrow 0$).

The operator (\ref{eq:T_gamma}) leads to a generator
operator of spatial translations corresponding
to a position-dependent linear momentum,
and consequently it represents a particle with position-dependent mass.
More generally, an Hermitian
generator operator of spatial translations was obtained in
\cite{costa-borges,mazharimousavi}, given by
\begin{eqnarray}
\label{eq:p_q}
	\hat{p}_q &=& \frac{(\hat{1} + \gamma_q \hat{x}) \hat{p}}{2}
				  + \frac{\hat{p}(\hat{1} + \gamma_q \hat{x})}{2}
				  \nonumber \\
			  &=& (\hat{1} + \gamma_q \hat{x})^{1/2} \hat{p}
				  (\hat{1} + \gamma_q \hat{x})^{1/2}.
\end{eqnarray}
A canonically conjugated space operator for the deformed
linear momentum operator is defined by
\begin{equation}
\label{eq:x_q}
	\hat{x}_q = \frac{\ln ( \hat{1} + \gamma_q \hat{x})}{\gamma_q}
			  = \xi \ln [\exp_q (\hat{x}/\xi)].	
\end{equation}
Hence, $(\hat{x}_q, \hat{p}_q)\rightarrow (\hat{x}, \hat{p})$
constitutes a point canonical transformation (PCT)
which maps a particle with constant mass
$m_0$ into another one with position-dependent mass.
In fact, the Hamiltonian operator
$
	\hat{K}(\hat{x}_q, \hat{p}_q) = \frac{1}{2m_0}\hat{p}_q^2 + \hat{V}(\hat{x}_q)
$
is mapped into $ \hat{H} (\hat{x}, \hat{p})= \hat{T} + \hat{V}(\hat{x})$
whose the kinetic energy operator is
\begin{equation}
\label{eq:general-hamiltonian-pdm}
	\hat{T} =
    \frac{1}{2}[m(\hat{x})]^{-1/4}\hat{p}\,[m(\hat{x})]^{-1/2}\hat{p}\,[m(\hat{x})]^{-1/4},
\end{equation}
with
\begin{equation}
\label{eq:m(x)}
m(x) = \frac{m_0}{(1 + \gamma_q x)^2}
\end{equation}
the effective mass, according to \cite{vonroos_1983}.
In consequence,  the time-dependent Schr\"odinger equation
for position-dependent mass in terms of wave function $\Psi(x, t)$ is
\begin{eqnarray}
\label{eq:schrodinger-equation-sho-pdm}
	i\hbar \frac{\partial \Psi (x,t)}{\partial t} &=&
	-\frac{{\hbar}^2 ( 1+\gamma_q x)^2}{2m_0} \frac{\partial^2\Psi (x,t)}{\partial x^2}
	\nonumber \\
	&&	
	-\frac{{\hbar}^2\gamma_q( 1+\gamma_q x)}{m_0} \frac{\partial \Psi (x,t)}{\partial x}
	\nonumber \\
	&&	
	-\frac{{\hbar}^2\gamma_{q}^2}{8m_0}\Psi (x,t)
	+ V(x) \Psi (x,t).
\end{eqnarray}
The Eq.~(\ref{eq:schrodinger-equation-sho-pdm}) can be adequately
rewritten by means of a field $\Phi_q(x, t)$ related to $\Psi (x, t)$ by
(see, for instance, \cite{Bravo-PRD-2016})
\begin{equation}
\label{eq:psi-phi-relation}
	\Psi (x, t) = \sqrt[4]{\frac{m(x)}{m_0}} \Phi_q (x, t)
				= \frac{\Phi_q (x, t)}{\sqrt{1+\gamma_q x}}.
\end{equation}
Thus, one obtains a $q$-deformed Schr\"odinger equation
\cite{costa-filho-2011,costa-filho-2013}:
\begin{equation}
\label{eq:deformed-schrodinger-equation}
	i\hbar \frac{\partial \Phi_q (x, t)}{\partial t} =
	-\frac{\hbar^2}{2m_0} \mathcal{D}_{\gamma_q}^2 \Phi_q (x, t)
	+ V(x) \Phi_q (x, t),
\end{equation}
where ${\mathcal{D}}_{\gamma_q} \equiv (1+\gamma_q x)\partial_x$
is a deformed derivative and
\begin{equation}
\label{eq:operator-H'}
\hat{H'}= \frac{\hbar^2}{2m_0}\mathcal{D}_{\gamma_q}^2
			+ V(\hat{x}).
\end{equation}
is a non-Hermitian operator.

Some remarks that deserve to be mentioned are the following.
First, one can see that the Eq.'s~(\ref{eq:schrodinger-equation-sho-pdm})
and (\ref{eq:deformed-schrodinger-equation}) represent systems
having the same energy spectrum
(isospectral systems). That is, although operator $\hat{H'}$
is non-Hermitian, it has real energy eigenvalues
(see \cite{RegoMonteiro-Nobre-2013-PRA} for some details).
Second, from Eq.~(\ref{eq:psi-phi-relation})
the probability densities $\rho(x,t) = |\Psi(x, t)|^2$ and
the $q$-deformed one $\varrho_q (x, t) = |\Phi_q (x, t)|^2$
satisfy,
\begin{equation}
\label{eq:rho_varrho}	
\rho (x, t) = \frac{\varrho_q(x, t)}{1 + \gamma_q x},
\end{equation}
and $\int_{x_i}^{x_f} \rho (x, t) dx =
	\int_{x_i}^{x_f} \varrho_q (x, t) d_q x =1.$
That is, while the distribution probability is normalized
by a standard integral, the $q$-deformed one is normalized by
a $q$-deformed one.

By last, we emphasize that there is an equivalence between
a Hermitian Hamiltonian system for position-dependent mass
and a deformed non-Hermitian one in terms of the $q$-derivative,
that results by replacing the field $\Psi(x,t)$
by the deformed one $\Phi_q(x, t)$.
This may be understood as the effect of the position-dependent mass
(\ref{eq:m(x)}) being imitated by a deformed derivative operator
in the Schr\"odinger equation (\ref{eq:schrodinger-equation-sho-pdm}).

\section{\label{sec:q-Bohm-theory}
		Bohmian quantum theory for $q$-deformed Schr\"odinger equation}

We present a deformed Bohmian theory for a position-dependent mass system,
and we obtain the dynamics in the classical limit.
Next, we explore a classical approach in the context of the Hamilton-Jacobi theory.

\subsection{$q$-Deformed Bohmian quantum theory}

In order to obtain a de Broglie--Bohm theory for the system
with position-dependent mass, we use the field $\Phi_q(x, t)$.
The same results can be obtained from the field $\Psi(x, t)$.
Consider the field of the $q$-deformed Schr\"odinger equation
expressed as a pilot wave, that is
$
	\Phi_q (x, t) = \sqrt{\varrho_q (x, t)}
	e^{iS_q(x, t)/ \hbar},
$
where $S_q(x, t)$ is a real phase that will
be physically interpreted in the following.
The $q$-deformed Schr\"odinger equation
(\ref{eq:deformed-schrodinger-equation}) leads us to
\begin{eqnarray}
\label{eq:real-imaginary}
	\displaystyle{ \left[
	\frac{1}{2m_0}(\mathcal{D}_{\gamma_q} S_q)^2
	+ V(x) - \frac{\hbar^2}{2m_0} \frac{1}{\sqrt{\varrho_q}}	
	\mathcal{D}_{\gamma_q}^2 \sqrt{\varrho_q}
	+ 	\frac{\partial S_q}{\partial t}
	\right] }  \nonumber \\
	\displaystyle{ + \frac{1}{2 \varrho_q} \left( \frac{i}{\hbar} \right)
	\left[ 	\frac{\partial \varrho_q}{\partial t} +
	\mathcal{D}_{\gamma_q} \left( \frac{\varrho_q}{m_0}
	\mathcal{D}_{\gamma_q} S_q \right) \right] = 0, }
\end{eqnarray}

From the imaginary part of Eq.~(\ref{eq:real-imaginary}),
we obtain a $q$-deformed continuity equation
\begin{equation}
\label{eq:eq:deformed-contituity-equation-pdm-system-S(x,t)}
	\frac{\partial \varrho_q (x,t)}{\partial t} +
	\mathcal{D}_{\gamma_q} \mathcal{J}_q(x, t) = 0,
\end{equation}
where the deformed current density $\mathcal{J}_q(x, t)$ is defined by
\begin{eqnarray}
\label{eq:currrent-density-and-S(x,t)}
	\mathcal{J}_q (x, t)
		&=& \frac{\varrho_q (x, t)}{m_0}
		\mathcal{D}_{\gamma_q} S_q(x, t)
 		\nonumber \\
		&=&
		\textrm{Re} \left[
		\Phi_q^{\ast} (x, t)
		\left( \frac{\hbar}{i}
		\mathcal{D}_{\gamma_q}\right)
		\left( \frac{\Phi_q	(x, t)}{m_0} \right)
		\right].
\end{eqnarray}
Equivalently, from Eq.~(\ref{eq:rho_varrho}) one can obtain
\begin{equation}
\label{eq:eq:deformed-contituity-equation-pdm-system-S(x,t)-2}
	\frac{\partial \rho (x,t)}{\partial t} +
	\frac{\partial J(x,t)}{\partial x} = 0,
\end{equation}
with $\mathcal{J}_q (x, t) = J(x, t)$
expressed in terms of $\rho(x,t)$ or $\Psi(x,t)$ as
\begin{eqnarray}
\label{eq:J(x,t)}
	J(x,t) &=&
		\frac{\rho (x, t)}{m(x)} \frac{\partial S_q}{\partial x}
		\nonumber \\
		&=&
		\textrm{Re}
		\left\{
		\Psi^{\ast} (x,t) \left( \frac{\hbar}{i}
		\frac{\partial}{\partial x} \right)
		\left[\frac{1}{m(x)} \Psi (x,t)\right]
		\right\}.
\end{eqnarray}
As in the standard case, the spatial variation
in the phase of the wave function is related to the flow of probability:
the more the phase changes,  the more intense is the probability flow.

From the real part of (\ref{eq:real-imaginary}),
we obtain the following $q$-deformed Hamilton-Jacobi equation
in the quantum formalism:
\begin{equation}
\label{eq:hamilton-jacobi}
	\frac{1}{2m_0} [\mathcal{D}_{\gamma_q} S_q (x, t)]^2
	+ V(x) + Q_q (x, t) + \frac{\partial S_q (x, t)}{\partial t} = 0,
\end{equation}
where, in terms of the mass function, we have
\begin{equation}
\label{eq:hamilton-jacobi2}
	\frac{1}{2m(x)} \left( \frac{\partial S_q (x, t)}{\partial x}\right)^2
	+ V(x) + Q_q (x, t) + \frac{\partial S_q (x, t)}{\partial t} = 0,
\end{equation}
with $Q_q (x, t)$ a deformed de Broglie-Bohm quantum potential given by
\begin{eqnarray}
\label{eq:Bohm-quantum-potential}
	Q_q (x, t) & \equiv &
		- \frac{\hbar^2}{2m_0} \frac{1}{|\Phi_q (x, t)|}	
		\mathcal{D}_{\gamma_q}^2 |\Phi_q (x, t)|
		\nonumber \\
		& = &
		- \frac{\hbar^2}{2m_0} \frac{1}{\sqrt{\varrho_q}}	
		\mathcal{D}_{\gamma_q}^2 \sqrt{\varrho_q }
 		\nonumber \\
		& = &
		\frac{\hbar^2}{2m_0} \left[
		\frac{1}{4} \left( \frac{1}{\varrho_q} \mathcal{D}_{\gamma_q} \varrho_q  \right)^2
		- \frac{1}{2\varrho_q}	\mathcal{D}_{\gamma_q}^2 \varrho_q \right].
\end{eqnarray}
From Eq.~(\ref{eq:rho_varrho}), the de Broglie-Bohm quantum potential
in terms of the density probability $\rho(x, t)$
(\ref{eq:Bohm-quantum-potential}) is
\begin{equation}
\label{eq:Q_q(x,t)}
Q_q(x, t) = Q_q^{(1)}(x, t) + Q_q^{(2)}(x, t) + Q_q^{(3)}(x, t),
\end{equation}
with
\begin{subequations}
\begin{eqnarray}
\label{eq:Q_q(x,t)-a}
	Q_q^{(1)} (x, t)
		& = &  \frac{\hbar^2}{2m(x)} \left[
		\frac{1}{4\rho^2} \left( \frac{\partial \rho}{\partial x} \right)^2
		- \frac{1}{2\rho}	\frac{\partial^2 \rho }{\partial x^2} \right],	\\
\label{eq:Q_q(x,t)-b}
	Q_q^{(2)} (x, t)
		& = & -\frac{\hbar^2}{4\rho} \frac{\partial \rho }{\partial x}
		\frac{d}{dx} \left[ \frac{1}{m(x)} \right],	\\
\label{eq:Q_q(x,t)-c}
	Q_q^{(3)} (x, t)
		&=&	-\frac{\hbar^2 \gamma_q^2}{8m_0},
\end{eqnarray}
\end{subequations}
in accordance to the Bohmian formulation proposed in
\cite{Plastino-Casas-Plastino-2001}.
The contributions $Q_q^{(1)}$ and $Q_q^{(2)}$ depend on the
probability density $\rho$, whereas $Q_q^{(3)}$ remains independent.

\subsection{Classical dynamics for system with position-dependent mass}

Taking the classical limit $\hbar \rightarrow 0$
in ~ (\ref{eq:hamilton-jacobi}), one obtains
\begin{equation}
\label{eq:classical-hamilton-jacobi}
	\frac{1}{2m_0} [\mathcal{D}_{\gamma_q} S_q(x, t)]^2
	+ V(x) + \frac{\partial S_q (x, t)}{\partial t} = 0,
\end{equation}
corresponding to a $q$-deformed Hamilton-Jacobi equation in classical mechanics.
Considering the separating of variables method, where $S_q(x, t) = W_q(x) - Et$,
and $W_q(x)$ is a $q$-deformed Hamilton's characteristic function, we have
\begin{eqnarray}
\label{eq:W(x)}
	W_q(x) &=& \pm \int^{x} \sqrt{2m(x') [E - V(x')]}dx' \nonumber \\
		 &=& \pm \int^{x} \sqrt{2m_0 [E - V(x')]}d_q x',
\end{eqnarray}
whose classical linear momentum is given by
\begin{equation}
\label{eq:momentum-S(x,t)}
p = \frac{\partial S_q(x, t)}{\partial x}
  =  \frac{dW_q(x)}{dx} = \sqrt{2m(x) [E - V(x)]}.
\end{equation}
It follows that the deformed action $S_q(x, t)$ can be written as
\begin{eqnarray}
\label{eq:S(x, t)-integral}
	S_q(x, t) &=& \pm \int^{x} \sqrt{2m(x') [E - V(x')]} dx' - Et \nonumber \\
			&=& \pm \int^{x} \sqrt{2m_0 [E - V(x')]} d_qx' - Et.
\end{eqnarray}
Then,
\begin{equation}
\label{eq:S(x,t)-action}
	\frac{dS_q(x,t)}{dt} = \frac{\partial S_q}{\partial x} \dot{x}
						 + \frac{\partial S_q}{\partial t}
					   = p \dot{x} - H = L,
\end{equation}
where we used (\ref{eq:momentum-S(x,t)})
and $H(x, \partial S/\partial x) + \partial S_q/\partial t = 0$.
It should be noted that $S_q(x, t)$ coincides with the classical action.
Thus, we have
$
	S_q = \int L(x, \dot{x})dt,
$	
with the Lagrangian function given by
\begin{equation}
\label{eq:lagrangian-pdm}
L(x, \dot{x}) = \frac{1}{2}m(x)\dot{x}^2 - V(x).
\end{equation}
Therefore, in the limit $\hbar \rightarrow  0$
the classical mechanics for position-dependent mass system is recovered.

\section{\label{sec:q-fisher}
		Fisher information for $q$-deformed Schr\"odinger equation}

Considering the $q$-deformed Schr\"odinger equation,
we apply the principle of minimum action to a deformed
FI, from which we obtain the complete Bohm quntum potential.
Kinetic energy operator for stationary states in the context
of the Thomas-Fermi-Dirac theory, along with Lagrangian and Hamiltonian
formulations are developed. Then, a deformed version of
the Cram\'{e}r-Rao bound is presented.

\subsection{$q$-Deformed Fisher information and Bohm quantum potential}

In a previous work \cite{Reginatto-1998}, Reginatto	derived
the Bohmian quantum theory for systems with constant mass
by mean of the principle of minimum Fisher information.
Plastino {\it et al.} \cite{Plastino-Casas-Plastino-2001}
extended the result for systems with
position-dependent mass whose kinetic energy operator is
$\hat{T'} =	\frac{1}{2} \hat{p} \frac{1}{m(\hat{x})} \hat{p}$,
which is different from the one given by Eq.~(\ref{eq:general-hamiltonian-pdm}).
The authors obtained directly the quantum potential
$Q_q^{(1)} + Q_q^{(2)}$ from the  Fisher functional given by
\begin{equation}
\label{eq:fisher-information-pdm}
I[ \rho ] = \int_{x_i}^{x_f} \frac{m_0}{m(x)} \frac{1}{\rho (x, t)}
			\left[ \frac{\partial \rho (x,t)}{\partial x} \right]^2 dx,
\end{equation}
through the functional derivative
\begin{equation}
\label{eq:PCP-bohm-potential}
	Q_q^{(1)} + Q_q^{(2)} = \frac{\hbar^2}{8m_0} \frac{\delta I}{\delta \rho}.
\end{equation}
From Eq.~(\ref{eq:fisher-information-pdm})
it can seen that for the constant mass case $m(x)= m_0$ one recovers the standard FI
denoted by $I_F[\rho]$.

Now, we propose the following deformed Fisher functional
\begin{eqnarray}
\label{eq:q-deformed-fisher-information}
	I_q[\varrho_q]
	&\equiv &
	\int_{x_i}^{x_f} \frac{1}{\varrho_q (x, t)}
   [\mathcal{D}_{\gamma_q} \varrho_q (x, t)]^2 d_q x,
\end{eqnarray}
where the ordinary derivative and integral operators are replaced
by the $q$-derivative and the $q$-integral respectively.
From the equivalence between the Schr\"odinger equation
for position-dependent mass
and the $q$-deformed Schr\"odinger equation,
one can relate the corresponding Fisher functionals
(the proposed  by Plastino and the deformed one).
Using the Eq.'s~(\ref{eq:rho_varrho})	
and (\ref{eq:q-deformed-fisher-information}), we have
that the $q$-deformed FI can be written as
\begin{eqnarray}
	I_q[{\varrho}_{q}] &=& \int_{x_i}^{x_f}
	\frac{1}{{\rho (x, t)}}
			     \left\{ \frac{\partial}{\partial x}
				 \left[ (1+\gamma_q x) \rho (x, t) \right] \right\}^2 dx
				 \nonumber \\
				 &=&  \int_{x_i}^{x_f} \gamma_q^2 \rho (x, t) dx
					 + \int_{x_i}^{x_f} 2\gamma_q (1 + \gamma_q x)
					   \frac{\partial \rho}{\partial x} dx
					 \nonumber \\
				 &&	+ \int \frac{(1+\gamma_q x)^2}{\rho(x,t)}
					  \left( \frac{\partial \rho}{\partial x} \right)^2 dx.
\end{eqnarray}
Thus, using Eq.~(\ref{eq:fisher-information-pdm})
and applying an integration by parts in the second term, this leads to
\begin{equation}
\label{eq:I_q-I}
	I_q[{\varrho_q}] = I[{\rho}] - \gamma_q^2.
\end{equation}
If we apply the functional derivative
\begin{equation}
		\frac{\delta I_q[\varrho_q]}{\delta \varrho_q}
		= \frac{\partial \mathcal{I}_q}{\partial \varrho_q}
		  - \mathcal{D}_{\gamma_q}
			 \left(
			\frac{\partial \mathcal{I}_q}{\partial (\mathcal{D}_{\gamma_q} \mathcal{I}_q)}
			\right),
\end{equation}
where
$\mathcal{I}_q (x, t) = \varrho_q(x, t)[\mathcal{D}_{\gamma_q} \ln \varrho_q(x,t)]^2$
is a $q$-deformed FI density, we obtain the {\it complete} deformed
de Broglie-Bohm quantum potential expressed by the functional derivative
[compare with Eq.~(\ref{eq:PCP-bohm-potential})]
\begin{equation}
\label{eq:q-Bohm-potential}
 	Q_q(x, t) = \frac{\hbar^2}{8m_0}
				\frac{\delta I_q[\varrho_q]}{\delta \varrho_q}.
\end{equation}
From Eq.'s~(\ref{eq:Q_q(x,t)}), (\ref{eq:Q_q(x,t)-c}), (\ref{eq:PCP-bohm-potential}),
(\ref{eq:I_q-I}) and (\ref{eq:q-Bohm-potential}), we have
\begin{equation}
 I_q[\varrho_q] - \frac{\delta I_q [\varrho_q]}{\delta \varrho_q} =
 I[\rho] - \frac{\delta I [\rho]}{\delta \rho},
\end{equation}
i.e., an invariant under transformation $\varrho_q \leftrightarrow \rho$.

\subsection{Kinetic energy operator for stationary states}

In terms of $\Psi(x, t)$, the Fisher functional
(\ref{eq:fisher-information-pdm}) can be written as
\begin{eqnarray}
\label{eq:fisher-information-psi}
	I[\rho] & = & 4\int_{x_i}^{x_f} \frac{m_0}{m(x)}
	\left( \frac{\partial \Psi^{\ast}}{\partial x} \right)
	\left( \frac{\partial \Psi}{\partial x} \right) dx
	\nonumber \\
	& &
	+\int_{x_i}^{x_f} \frac{m_0}{m(x)} |\Psi (x, t)|^2
	\left( \frac{1}{\Psi}\frac{\partial \Psi}{\partial x}
	-\frac{1}{\Psi^{\ast}} \frac{\partial \Psi^{\ast}}{\partial x} \right)^{2}.
\end{eqnarray}
One has that $\hat{p} = -i\hbar \partial_x$ is the momentum operator
in the representation $\{| \hat{x} \rangle\}$,
while from $\Psi(x, t) = \sqrt{\rho(x, t)} e^{iS_q(x, t)/\hbar}$
and Eq.~(\ref{eq:momentum-S(x,t)}),
the classical linear momentum obeys the relation
$
p = \frac{\hbar}{2i} \left( \frac{1}{\Psi}\frac{\partial \Psi}{\partial x}
    -\frac{1}{\Psi^{\ast}} \frac{\partial \Psi^{\ast}}{\partial x} \right).
$
Therefore, it follows that the Fisher functional
(\ref{eq:fisher-information-psi}) can be interpreted
as a  measure of nonclassicality between
the quantum kinetic term and the classic one given by
\begin{equation}
\label{eq:nonclassicality}
I[{\rho}] = \frac{4m_0}{\hbar^2} \left(
\left\langle \hat{p} \frac{1}{m(\hat{x})} \hat{p} \right\rangle
- \left\langle \frac{p^2}{m(x)} \right\rangle_{\textrm{classic}}
\right).
\end{equation}
It should be noted that (\ref{eq:nonclassicality})
has been also studied for a constant mass in \cite{Hall-PRA-2000}.

For stationary states we have a constant current density $J$,
so using the Eq.~(\ref{eq:J(x,t)}) and integrating by parts,
the classical contribution to $I[\rho]$ results
\begin{eqnarray}
\left\langle \frac{p^2}{m(x)} \right\rangle_{\textrm{classic}}
&=& \int_{x_i}^{x_f}\frac{\rho(x,t)}{m(x)}
\left( \frac{\partial S_q}{\partial x} \right)^2 dx
\nonumber \\
&=& - \int_{x_i}^{x_f} S_q(x, t)  \frac{\partial}{\partial x} \left[
\frac{\rho(x,t)}{m(x)}\left( \frac{\partial S_q}{\partial x}
\right) \right] dx
\nonumber \\
&=& 0.
\end{eqnarray}
Accordingly, the FI for the stationary states $\psi_n(x)$ is
\begin{equation}
\label{eq:I-stationary-states}
	I[ \rho_n ] = 4\int_{x_i}^{x_f} \frac{m_0}{m(x)}
				  \left[\frac{d\psi_n (x)}{dx} \right]^{2} dx
				= \frac{8m_0}{\hbar^2}\langle \hat{T'} \rangle.
\end{equation}

Considering the $q$-deformed FI, we have for the stationary states
\begin{equation}
\label{eq:I_q-stationary-states}
	I_q[ \varrho_{q,n} ]
	= 4\int_{x_i}^{x_f} [\mathcal{D}_{\gamma_q} \varphi_{q,n} (x)]^2 d_q x
	= \frac{8m_0}{\hbar^2}\langle \hat{T}\rangle,
\end{equation}
in accordance to the von Weizs\"acker's kinetic energy functional operator
\cite{weizsacker-1935} in Thomas-Fermi-Dirac theory.

Some researchers
\cite{Falaye-2016,YanezNavarro-2014,Macedo-Guedes-2015}
have considered the information theory for systems with
position-dependent mass by means using the standard FI
\begin{equation}
\label{eq:standard-Fisher}
I_F [\rho ] = 4\int_{x_i}^{x_f} \left[\frac{d\psi_n (x)}{dx} \right]^{2} dx
			= \frac{\langle \hat{p}^2 \rangle}{(\hbar/2)^2}.
\end{equation}
Section \ref{sec:appl} uses Eq.~(\ref{eq:standard-Fisher}) for completeness.

\subsection{$q$-Deformed Lagrangian and
				Hamiltonian formulations from variational principle}

Consider the Lagrangian formulation by defining
the deformed Lagrangian density as
\begin{eqnarray}
 \label{eq:density-lagrangian_L_q}
 {\mathcal{L}}_q (x, t)
	& = &
	\left\{ \frac{\partial S_q (x, t)}{\partial t}
	+ \frac{[\mathcal{D}_{\gamma_q} S_q(x, t)]^2}{2m_0}
	+ V(x) \right\} \varrho_q (x, t)
	\nonumber \\
	&& + \frac{\hbar^2}{8m_0} \frac{1}{\varrho_q (x, t)}
					     [\mathcal{D}_{\gamma_q} \varrho_q (x, t)]^2,
\end{eqnarray}
and its corresponding action as
\begin{eqnarray}
\label{eq:density-lagrangian_L_q-2}
  A	&=& \int_{t_i}^{t_f} \int_{x_i}^{x_f}
	\mathcal{L}_q (x, t)  d_q x dt
	\nonumber \\
	&=& \int_{t_i}^{t_f} \int_{x_i}^{x_f}
	\left[ \frac{\partial S_q}{\partial t} +  \frac{(\mathcal{D}_{\gamma_q} S_q)^2}{2m_0}
	+ V  \right] \varrho_q {d_{q}x}{dt}
	\nonumber \\
	&& + \frac{\hbar^2}{8m_0} \int_{t_i}^{t_f} I_q[\varrho_q] dt,
\end{eqnarray}
then it follows that by applying the variational principle $\delta A = 0$ we get
the equations of motion
(\ref{eq:eq:deformed-contituity-equation-pdm-system-S(x,t)})
and (\ref{eq:hamilton-jacobi}) for the fields $S_q(x, t)$
and $\varrho_q(x, t)$ related to the Bohmian quantum formalism.

Alternatively, the equations of motion
(\ref{eq:eq:deformed-contituity-equation-pdm-system-S(x,t)-2})
and (\ref{eq:hamilton-jacobi2}),
associated to the Schr\"odinger equation for systems with position dependent mass
(\ref{eq:schrodinger-equation-sho-pdm})
emerge from the variational principle applied to the standard Lagrangian density
\begin{eqnarray}
 \label{eq:density-lagrangian_L-pdm}
 {\mathcal{L} (x, t)} & = &
	 \left[ \frac{\partial S_q (x, t)}{\partial t}
	+ \frac{1}{2m(x)} \left( \frac{\partial S_q (x, t)}{\partial x} \right)^{2}
	+ V(x)  \right] \rho (x, t)
	\nonumber \\
	&&   + \frac{\hbar^2}{8 m(x)} \frac{1}{\rho (x, t)}
					     \left[ \frac{ \partial \rho (x, t)}{\partial x} \right]^2,
\end{eqnarray}
whose corresponding action is
\begin{eqnarray}
A   &=& \int_{t_i}^{t_f} \int_{x_i}^{x_f}
	\mathcal{L} (x, t)  dx dt
	\nonumber \\
	&=& \int_{t_i}^{t_f} \int_{x_i}^{x_f}
	\left[ \frac{\partial S_q}{\partial t} +  \frac{1}{2m(x)}
			\left( \frac{\partial S_q}{\partial x} \right)^2
	+ V  -\frac{\hbar^2\gamma_q^2}{8m_0} \right] \rho {dx}{dt}
	\nonumber \\
	&& + \frac{\hbar^2}{8m_0} \int_{t_i}^{t_f} I[\rho] dt.
\end{eqnarray}

A Hamiltonian formulation also can be developed.
For the Hamiltonian (\ref{eq:operator-H'}) we have that
the energy of the system is given by the $q$-integral
\begin{eqnarray}
\label{eq:Hamiltonian-density}
		E & = &
   	    \int_{x_i}^{x_f} \Phi_q^{\ast}(x, t)  \hat{H'} \Phi_q (x, t) d_q x
	    \nonumber \\
       &=& 	\int_{x_i}^{x_f} \Phi_q^{\ast} (x, t) \left[
			-\frac{\hbar^2}{2m_0}\mathcal{D}_{\gamma_q}^{2}
			+ V(x) \right] \Phi_q (x, t) d_q x.
\end{eqnarray}
Applying an integration by parts, we obtain
\begin{equation}
		E = \int_{x_i}^{x_f}
				\left\{  \frac{\hbar^2}{2m_0}|\mathcal{D}_{\gamma_q} \Phi_q (x, t)|^2
				+V(x)  |\Phi_q (x, t)|^2
				\right\} d_q x,
\end{equation}
and then, in terms of $S_q(x, t)$ and $\varrho_q(x, t)$, we can write
\begin{eqnarray}
\label{eq:energy-q-density}
		E & = & \int_{x_i}^{x_f} \mathcal{H}_q d_q x
		\nonumber \\
		  &=&  \int_{x_i}^{x_f} \left[
			   \frac{(\mathcal{D}_{\gamma_q} S_q)^2}{2m_0}
			   + V(x)  \right] \varrho_q d_q x
			   + \frac{\hbar^2}{8m_0} I_q[{\varrho_q}],
\end{eqnarray}
where $\mathcal{H}_q$ is a deformed Hamiltonian density.

From the Hamilton's equations for the fields $\varrho_q(x, t)$
and $S_q(x, t)$, we recover the equations of motion:
\begin{eqnarray}
\frac{\partial \varrho_q}{\partial t}
		&=& \frac{\delta \mathcal{H}_q}{\delta S_q}
		= \frac{\partial \mathcal{H}_q}{\partial S_q}
		- \mathcal{D}_{\gamma_q} \left[
		\frac{\partial \mathcal{H}_q}{\partial (\mathcal{D}_{\gamma_q} S_q)}
		\right]
		\nonumber \\
		&=& -\mathcal{D}_{\gamma_q}
			\left(
			\frac{1}{m_0} \varrho_q \mathcal{D}_{\gamma_q}  S_q
			\right),
\end{eqnarray}
and
\begin{eqnarray}
\frac{\partial S_q}{\partial t}
		&=& -\frac{\delta \mathcal{H}_q}{\delta \varrho_q}
		= -\frac{\partial \mathcal{H}_q}{\partial \varrho_q}
		+ \mathcal{D}_{\gamma_q} \left[
		\frac{\partial \mathcal{H}_q}{\partial (\mathcal{D}_{\gamma_q} \varrho_q)}
		\right]
		\nonumber \\
		&=& -\frac{(\mathcal{D}_{\gamma_q} S_q)^2}{2m_0} - V - Q_q.
\end{eqnarray}

Similarly, there is an equivalent Hamiltonian formulation corresponding
to the Schr\"odinger equation for system position-dependent mass.
In this case, the energy of the system is given by
\begin{eqnarray}
\label{eq:Hamiltonian-density-2}
 		E &=&
   	    \int_{x_i}^{x_f}
		\Psi^{\ast}(x, t)  \hat{H} \Psi (x, t) d x \nonumber \\
	  &=& \int_{x_i}^{x_f}
		  \frac{\hbar^2}{2m(x)}\left|\frac{\partial \Psi (x, t)}{\partial x}\right|^2 dx
	  \nonumber \\
	   && + \int_{x_i}^{x_f}
			\left( V(x) - \frac{\hbar^2 \gamma_q^2}{8m_0} \right) |\Psi (x, t)|^2 dx,
\end{eqnarray}
which in terms of $\rho(x, t)$ and $S_q(x, t)$ can be recasted as
\begin{eqnarray}
\label{eq:energy-density-m(x)}
		E & = & \int_{x_i}^{x_f} \mathcal{H} d x \nonumber \\
		  & = &  \int_{x_i}^{x_f}
					\left[ \frac{1}{2m(x)} \left(\frac{\partial S_q}{\partial x}\right)^2
						+ V - \frac{\hbar^2 \gamma_q^2}{8m_0}  \right] \rho  dx
						+ \frac{\hbar^2}{8m_0} I[{\rho}],
\end{eqnarray}
where $\mathcal{H}$ is the Hamiltonian density for the position-dependent mass system.

Again, from the Hamilton's equations for the fields
$\rho(x, t)$ and $S_q(x, t)$ we recover the equations of motion
\begin{eqnarray}
\frac{\partial \rho}{\partial t}
		&=& \frac{\delta \mathcal{H}}{\delta S_q}
		= \frac{\partial \mathcal{H}}{\partial S_q}
		- \frac{\partial}{\partial x}
		\left[
		\frac{\partial \mathcal{H}}{\partial (\partial_x S_q)}
		\right]
		\nonumber \\
		&=& -\frac{\partial}{\partial x}
			\left(
			\frac{\rho (x,t) }{m(x)}\frac{\partial S_q}{\partial x}
			\right),
\end{eqnarray}
and
\begin{eqnarray}
\frac{\partial S_q}{\partial t}
		&=& -\frac{\delta \mathcal{H}}{\delta \rho}
		= -\frac{\partial \mathcal{H}}{\partial \rho}
		+ \frac{\partial}{\partial x} \left[
		\frac{\partial \mathcal{H}}{\partial (\partial_x \rho)}
		\right]
		\nonumber \\
		&=& -\frac{1}{2m(x)} \left( \frac{\partial S_q}{\partial x}\right)^2
			- V - Q_q.
\end{eqnarray}

\subsection{$q$-Deformed Cram\'er-Rao bound}

The Cram\'er-Rao bound states an uncertainty principle for
probability distributions in terms of the FI and the variance.
If $\rho(x)$ is a probability distribution and
$(\Delta x)^2=\int \rho(x) (x - \langle x \rangle)^2 dx$
is the variance, then this is given by
$
	I_F[\rho] (\Delta x)^2 \geq 1.
$
It can be shown that the family of distributions
that minimize the Cram\'er-Rao bound are the Gaussian ones.

Following a way similar proposed by Furuichi \cite{Furuichi-JMP},
now we develop a Cram\'er-Rao bound associated with the proposed $q$-deformed FI
(\ref{eq:q-deformed-fisher-information}).
Consider the expected value in terms of the $q$-deformed
density probability for stationary states
$
	\langle f(\hat{x}) \rangle = \int_{x_i}^{x_f} f(x) {\varrho}_{q} (x) d_q x
	 = \int_{x_i}^{x_f} f(x) {\rho}(x) dx.
$
In particular, we have that
$
	I_q[\varrho_q] = \langle \Omega_q^2 \rangle,
$
where $\Omega_q = \mathcal{D}_{\gamma_q} \ln \varrho_q$
is a $q$-deformed score function.
Thus, one has
\begin{eqnarray}
	\langle (x - \langle \hat{x} \rangle) \Omega_q (x) \rangle
		&=&
		\int_{x_i}^{x_f} (x - \langle \hat{x} \rangle) \varrho_q(x)  \Omega_q(x) d_q x
		\nonumber \\
		&=& -\int_{x_i}^{x_f} \varrho_{q}(x) dx
		\nonumber \\
		&=& -(1 + \gamma_q \langle \hat{x} \rangle) \nonumber
\end{eqnarray}
Therefore, it follows that
\begin{eqnarray}
	0 	& \leq & \left\langle \left[
	\Omega_q(x) + \frac{(x - \langle \hat{x} \rangle)}{(\Delta x)^2}
	\right]^2 \right\rangle
	\nonumber \\
	&=& I_q [\varrho_q] + \frac{2}{(\Delta x)^2}
					  \langle (x - \langle \hat{x} \rangle) \Omega_q(x) \rangle
		+ \frac{\langle (x - \langle \hat{x} \rangle)^2 \rangle}{(\Delta x)^4}
	\nonumber \\
	&=& I_q [\varrho_q] -\frac{2(1 + \gamma_q \langle \hat{x} \rangle)}{(\Delta x)^2}
						+ \frac{1}{(\Delta x)^2} \nonumber
\end{eqnarray}
from which one obtains
\begin{equation}
\label{eq:q-Cramer-Rao}
	I_q[\varrho_q] (\Delta x)^2 \geq 1 + 2\gamma_q \langle \hat{x} \rangle.
\end{equation}
that constitutes a $q$-deformed version of the Cram\'er-Rao bound.
Note that since $I_q[\varrho_q] \rightarrow I_F[\varrho]$
when $\gamma_q \rightarrow 0$, then the standard one is recovered
in the limit $\gamma_q \rightarrow 0$.

\section{\label{sec:appl}
		Application: particle in an infinite square potential well}

Consider a particle with position-dependent mass $m(x)$ given by
Eq.~(\ref{eq:m(x)}) in an infinite one-dimensional square potential well
of width $L$. The eigenfunctions for this problem are given by \cite{costa-borges,Schmidt-2006}
\begin{equation}
 \label{eq:psi_n}
   \psi_{n}(x) = \frac{A_{q}}{\sqrt{1+ \gamma_q x}}
                 \sin \left[ \frac{k_{q,n}}{\gamma_q}\ln (1+ \gamma_q x) \right],
\end{equation}
for $0 \leq x \leq L$ and $\psi_n(x) = 0$ otherwise,
where $A_q^2 = 2/L_q$, $k_{q,n} = n \pi/L_q$ ($n$ is a non integer), and
$
L_q = \gamma_q^{-1} \ln (1+\gamma_q L)
$
is the length of the box at the deformed space $\{ |\hat{x}_q\rangle \}$
obtained by the transformation (\ref{eq:x_q}).
The corresponding solutions using the $q$-deformed Schr\"odinger equation are
\begin{equation}
\label{eq:phi_n}
   \varphi_{q,n}(x) = A_q \sin \left[ \frac{k_{q,n}}{\gamma_q}
					  \ln (1 + \gamma_q x) \right],
\end{equation}
for $0 \leq x \leq L$ and $\varphi_{q,n}(x) = 0$ otherwise.
In this case, the expected value of the kinetic energy operator
coincides with the energy of the eigenstates, given by
\begin{equation}
\label{eq:E_n-particle-in-well}
			 E_n = \langle \hat{T} \rangle
				 = \frac{\hbar^2 \pi^2  \gamma_q^2 n^2}{2m_0 \ln^2 (1 + \gamma_q L )}.
\end{equation}
The expected values for
$ \langle \hat{x} \rangle $,
$ \langle \hat{x}^2 \rangle $,
$ \langle \hat{p} \rangle $ and
$ \langle \hat{p}^2 \rangle $
are \cite{costa-borges}
\begin{subequations}
\label{eq:expectation_values_quantum}
\begin{eqnarray}
  \label{eq:x-med-quantum}
  \langle \hat{x} \rangle &=&
   \frac{\gamma_q L - \ln(1 + \gamma_q L )}{\gamma_q \ln(1 + \gamma_q L )}
	-\frac{L \ln(1 + \gamma_q L )}{\ln^2(1 + \gamma_q L ) +(2\pi n)^2},
\end{eqnarray}
\begin{eqnarray}
 \label{eq:x^2-med-quantum}
 \langle \hat{x}^2 \rangle &=&
 \frac{\gamma_q^2 L^2 - 2\gamma_q L + 2\ln(1 + \gamma_q L )}{2\gamma_q^2 \ln (1 + \gamma_q L )}
 \nonumber \\
 &&
 +\frac{[1-(1+\gamma_q L)^2]\ln(1 + \gamma_q L )}{2\gamma_q ^2 [\ln^2 (1 + \gamma_q L ) + n^2{\pi}^2]}
 \nonumber \\
 &&+\frac{2\gamma_q L \ln(1 + \gamma_q L )}{\gamma_q ^2 [\ln^2 (1 + \gamma_q L ) + 4n^2{\pi}^2]},
\end{eqnarray}
\begin{equation}
\label{eq:value-expected-p}
 \langle \hat{p} \rangle = 0,
\end{equation}
\begin{equation}
\label{eq:expected-value-p^2}
 \langle \hat{p}^2 \rangle =
 \frac{\hbar^2 k_{q,n}^2 [(1+\gamma_q L)^2 - 1]}{2(1+\gamma_q L)^2\ln (1 + \gamma_q L)}
 \left[ 1 + \frac{{\gamma}_q^2}{4({k_{q,n}^2 + {\gamma}_q^2})} \right],
\end{equation}
\end{subequations}
which satisfy the uncertainty principle
$\Delta x \Delta p \geq \hbar/2$ \cite{costa-borges}.

Now, in order to analyze this result we use the FI.
The $q$-deformed FI for the stationary states  (\ref{eq:phi_n}) is
\begin{eqnarray}
\label{eq:I_q-well}
 I_q[ \varrho_q ] &=& 4A_q^2 k_{q,n}^2 \int_{0}^{L}
			  \cos^2 \left[ \frac{k_{q,n}}{\gamma_q} \ln (1+ \gamma_q x) \right]
			\frac{dx}{1+\gamma_q x} \nonumber \\	
 		  &=& 4k_{q,n}^2,
\end{eqnarray}
in accordance with the Eq.~(\ref{eq:I_q-stationary-states})
for the kinetic energy (\ref{eq:E_n-particle-in-well}).
Note that Eq.~(\ref{eq:I_q-well}) has the same form than the
constant mass case, used in \cite{Lopez-Rosa-2011}.
From Eq.'s~(\ref{eq:I_q-I}) and (\ref{eq:I_q-well}),
the FI proposed by Plastino {\it et al.} results
$
	I[\rho] = 4k_{q,n}^2 + \gamma_q^2.
$
We can also calculate the standard FI
disregarding the effect of mass locality on functional.
From Eq.~(\ref{eq:standard-Fisher}) and (\ref{eq:expected-value-p^2}), we get
\begin{equation}
I_F [\rho] = \frac{2 k_{q,n}^2[(1 + \gamma_q L)^2 - 1]}
			  {(1 + \gamma_q L)^2 \ln ( 1 + \gamma_q L)}
			  \left[ 1 + \frac{\gamma_q^2}{4(\gamma_q^2 + k_{q,n}^2)} \right].
\end{equation}

\begin{figure}[hbt]
\centering
\begin{minipage}[b]{0.75\linewidth}
\includegraphics[width=\linewidth]{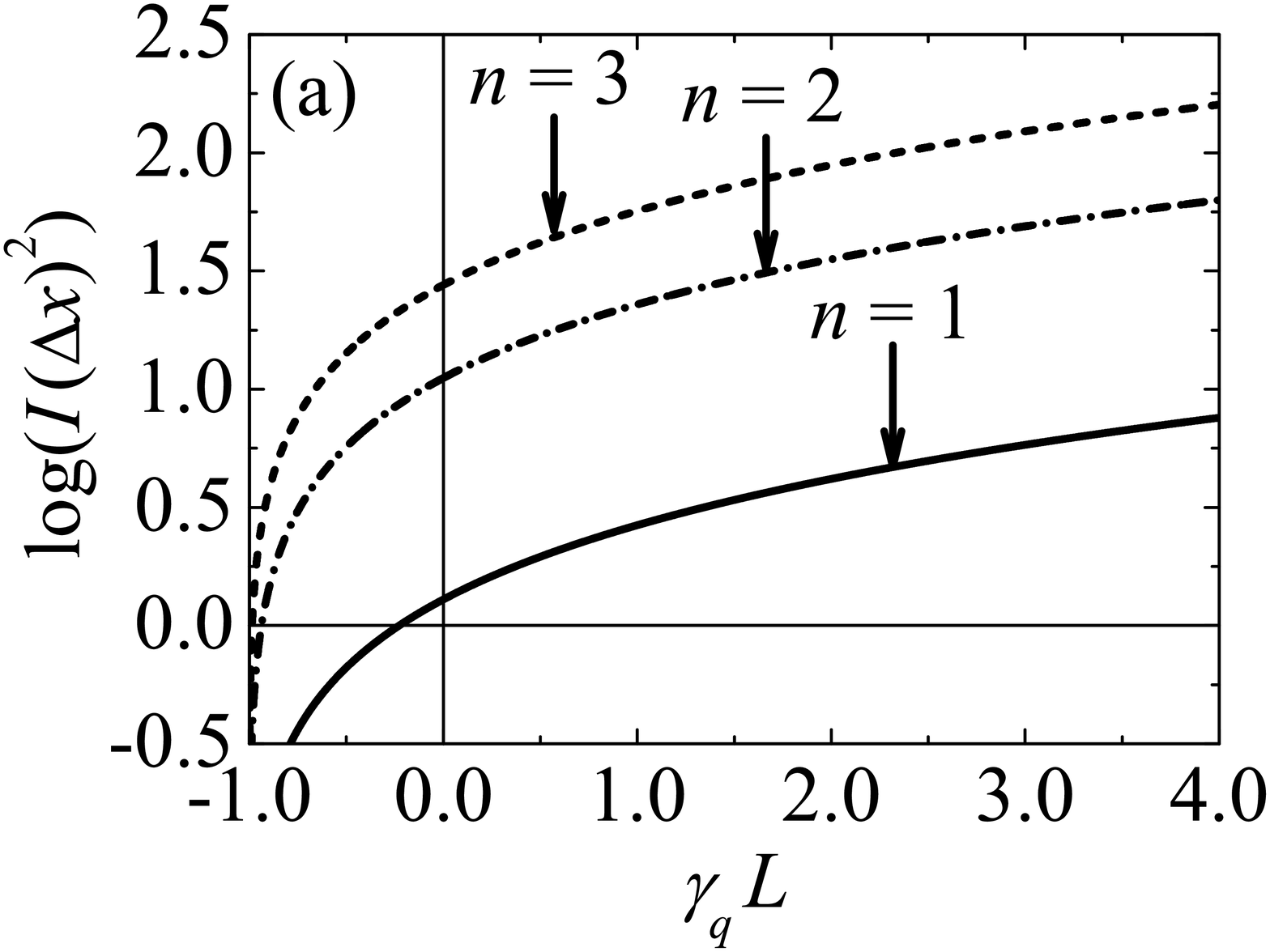}
\end{minipage}\\
\begin{minipage}[b]{0.75\linewidth}
\includegraphics[width=\linewidth]{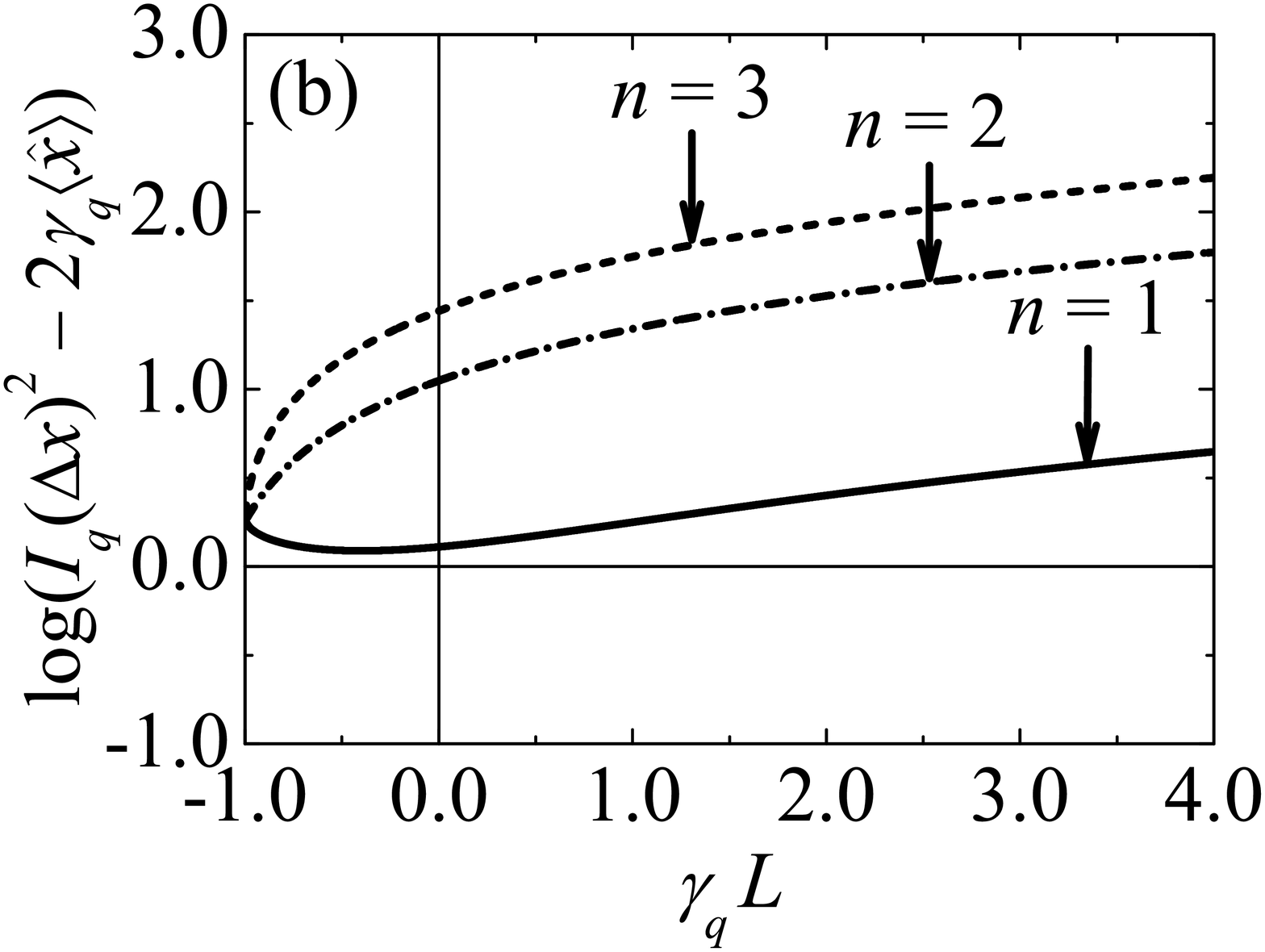}
\end{minipage}\\
\begin{minipage}[b]{0.75\linewidth}
\includegraphics[width=\linewidth]{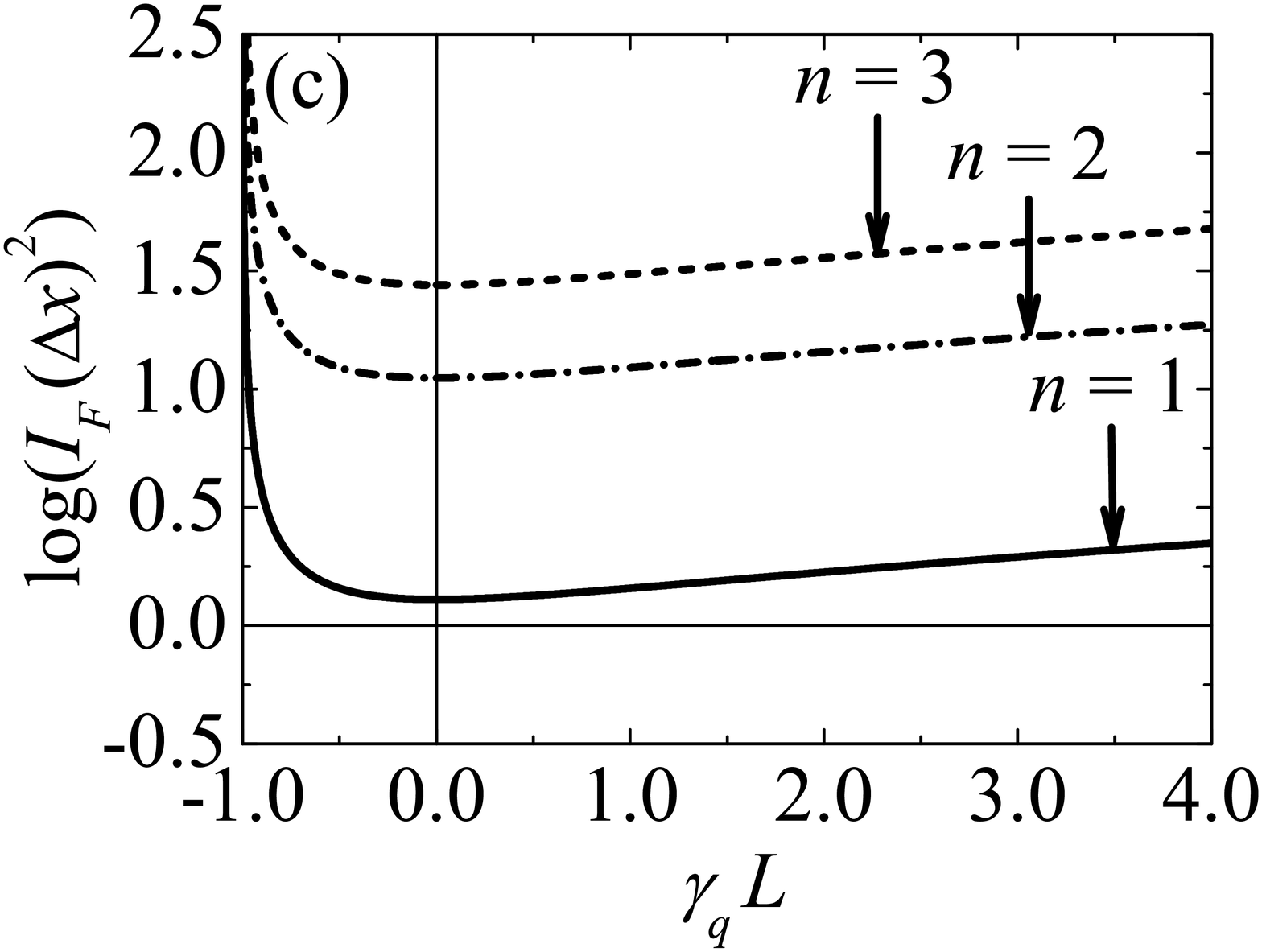}
\end{minipage}
\caption{\label{fig:1}
Logarithms of
(a) $I_q (\Delta x)^2$,
(b) $q$-deformed Cram\'er-Rao inequality (\ref{eq:q-Cramer-Rao}) and
(c) $I_F (\Delta x)^2$
for the three lower excited states of a position-dependent mass particle
in an infinite square well as a function of  $\gamma_qL$.
}
\end{figure}

Figure \ref{fig:1} (a) shows the relation $I (\Delta x)^2$
as function of $\gamma_q L$. Note that $I (\Delta x)^2 < 1$
as $\gamma_q L$ approaches to $-1$.
Figure \ref{fig:1} (b) shows that the $q$-deformed
Cram\'er-Rao inequality (\ref{eq:q-Cramer-Rao})
is satisfied for different values $\gamma_qL$.
Figure \ref{fig:1} (c) shows the standard
Cram\'er-Rao  inequality $I_F(\Delta x)^2$.
One can see that the Cram\'er-Rao inequality is only satisfied for their
standard version and the $q$-deformed one, thus showing the
consistence of the $q$-deformed structure.
In all cases,
the Cram\'er-Rao inequality
for particle with constant mass is recovered
in the limit $\gamma_q L \rightarrow 0$.

\section{\label{sec:conclusion}
		Conclusions}

\begin{figure}[hbt]
\centering
\includegraphics[width=\linewidth]{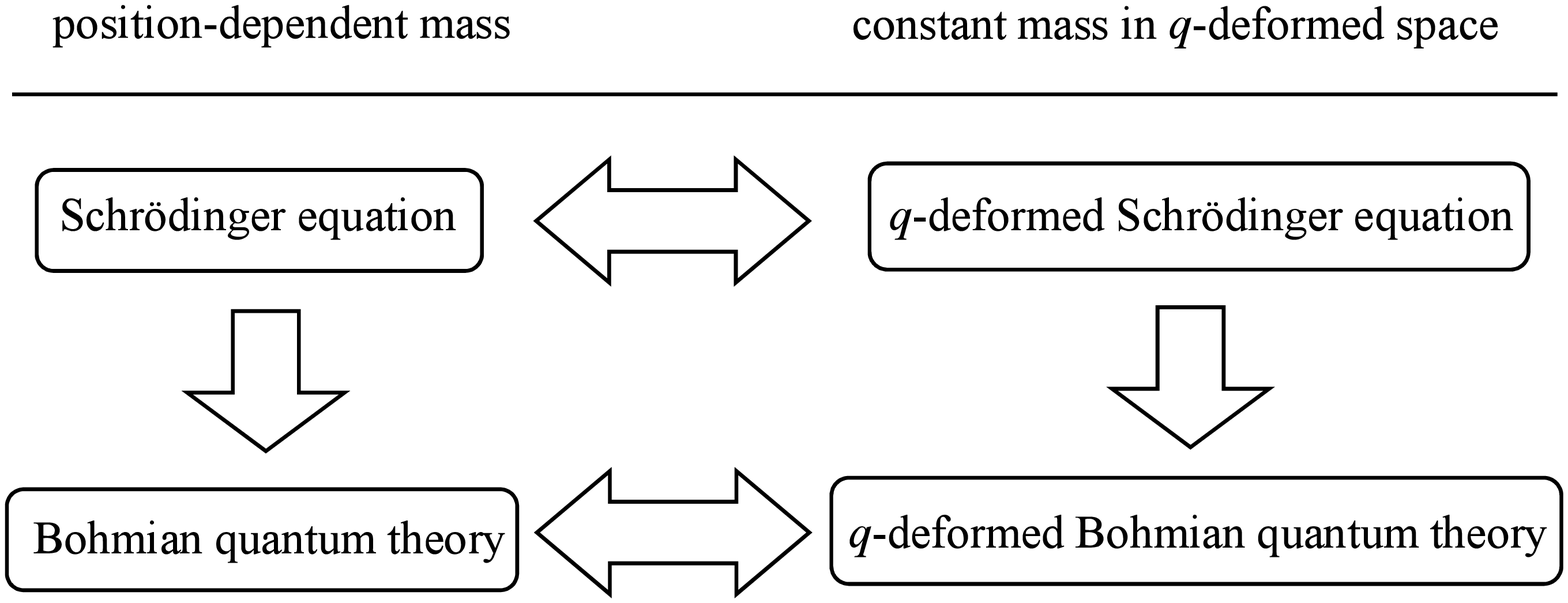}
\caption{\label{fig:2}
The standard treatment for position-dependent mass systems
is equivalent to the $q$-deformed one where the space
is deformed instead of the mass, that remains constant.
}
\end{figure}

We have proposed an alternative way for obtaining the Bohmian quantum formalism using
the deformed Schr\"odinger equation and the de Broglie wave pilot interpretation,
in the context of a $q$-algebra structure.
Specifically, a deformed derivative was used to represent a particle with a
position-dependent mass with the advantage of controlling the deformation
by means of the parameter $\gamma_q$.
Additionally, we have proposed a deformed Fisher functional that
that allows one to derive a deformed Hamilton-Jacobi equation
which emerges from the Bohmian formalism
for a system with a position-dependent effective mass.
The Lagrangian and Hamiltonian formulations have been also established.

Then, we have formulated a deformed Cram\'er-Rao bound associated
with the deformed Fisher functional proposed.
We have illustrated with a particle confined in a infinite square potential well.
We found that the deformed Cram\'er-Rao bound is satisfied by the stationary states,
while the product of the FI for position-dependent mass
(previously defined in \cite{Plastino-Casas-Plastino-2001})
with the variance of the position
violates the inequality, i.e., it results lower than one.
This can be interpreted as an evidence that the $q$-structure is preserved,
at least in the example studied. By the same way, another potentials
could be analyzed, e.g., the harmonic oscillator, Morse and Coulomb potentials, etc.
Furthermore, one can see that the scenario with position-dependent
mass can be treated, equivalently, by means of a $q$-deformed one where
the mass remains constant at the expense of deforming the space. An scheme
is shown in Fig. 2.

Finally, we mention that the deformed FI
and the Cram\'er-Rao bound presented in this work are inspired
in the $q$-deformed Schr\"odinger equation proposed in \cite{costa-filho-2011}.
Thus, they are not exhaustive
and further generalizations based on other deformed derivatives
could be formulated in future researches.

\section*{Acknowledgments}
This work was partially supported by National
Institute of Science and Technology for Complex Systems (INCT-SC) and Capes.




\begin{thebibliography}{99}

\bibitem{jaynes_1957}
E. T. Jaynes,
Phys. Rev. 106 (4) (1957) 620; 108 (2) (1957) 171.


\bibitem{Frieden-book}
B. R. Frieden,
Science from Fisher information: a unification,
Cambridge University Press, Cambridge, England, 2004.


\bibitem{Reginatto-1998}
M. Reginatto,
Phys. Rev. A 58 (3) (1998)  1775.


\bibitem{Hall-Reginatto-2002}
M. J. W. Hall, M. Reginatto,
J. Phys. A: Math. Gen. 35 (2002) 3289.


\bibitem{Bohm-1952}
D. Bohm,
Phys. Rev. 85 (1952) 166.


\bibitem{deBroglie-1927}
L. de Broglie,
J. Phys. Radium 8 (1927) 225.


\bibitem{Plastino-Casas-Plastino-2001}
A. R. Plastino, M. Casas, A. Plastino,
Phys. Lett. A 281 (2001) 297.


\bibitem{vonroos_1983}
O. von Roos,
Phys. Rev. B 27 (12) (1983) 7547.


\bibitem{Serra-Lipparini-1997}
L. Serra, E. Lipparini,
Europhys. Lett. 40 (6) (1997) 667.


\bibitem{Bencheikh-et-al-2004}
K. Bencheikh, K. Berkane, S. Bouizane,
J. Phys. A: Math. Gen. 37 (45) (2004)  10719.


\bibitem{Ioffe-2016}
M. V. Ioffe, E.V. Kolevatova, D.N. Nishnianidze,
Phys. Lett. A 380 (2016) 3349.


\bibitem{Ranada-2016}
Manuel F. Ra\~nada,
Phys. Lett. A 380 (2016) 2204.


\bibitem{Barranco-1997}
M. Barranco, M. Pi, S. M. Gatica, E. S. Hern\'andez, J. Navarro,
Phys. Rev. B 56 (1997) 8997.


\bibitem{Aquino_1998}
N. Aquino, G. Campoy, H. Yee-Madeira,
Chem. Phys. Lett. 296 (1998) 111.


\bibitem{Richstone_1982}
D. O. Richstone, M. D. Potter,
Astrophys. J. 254 (1982) 451.


\bibitem{Li-Guo-Jiang-Hu}
K. Li, K. Guo, X. Jiang, M. Hu,
Optik-International Journal for Light and Electron Optics 132 (2017) 375.


\bibitem{Alhaidari-2004}
A. D. Alhaidari,
Phys. Lett. A 322 (2004) 72.


\bibitem{Alimohammadi-Hassanabadi-Zare-2017}
M. Alimohammadi, H. Hassanabadi, S. Zare,
Nucl. Phys. A 960 (2017) 78.


\bibitem{BenDaniel-Duke-1966}
D. J. BenDaniel, C. B. Duke,
Phys. Rev. 152 (1966)  683.


\bibitem{Gora-Williams-1969}
T. Gora, F. Williams,
Phys. Rev. 177 (1969) 1179.


\bibitem{Zhu-Kroemer-1983}
Q. G. Zhu, H. Kroemer,
Phys. Rev. B 27 (6) (1983) 3519.


\bibitem{Li-Kuhn-1993}
T. L. Li, K. J. Kuhn,
Phys. Rev. B 47 (1993) 12760.


\bibitem{costa-filho-2011}
R. N. Costa Filho, M. P. Almeida, G. A. Farias, J. S. Andrade Jr.,
Phys. Rev. A 84 (2011) 050102(R).


\bibitem{costa-filho-2013}
R. N. Costa Filho, G. Alencar, B.-S. Skagerstam, J. S. Andrade Jr,
Europhys. Lett. 101 (2013) 10009.


\bibitem{mazharimousavi}
S. H. Mazharimousavi,
Phys. Rev. A 85 (2012) 034102;
89 (2014) 049904(E) [erratum].


\bibitem{costa-borges}
B. G. da Costa, E. P. Borges,
J. Math. Phys. 55 (2014) 062105.


\bibitem{borges_2004}
E. P. Borges,
Physica A 340 (2004) 95.


\bibitem{Entropia-Tsallis}
C. Tsallis,
J. Stat. Phys. 52 (1988) 479.


\bibitem{Tsallis-book-2009}
C. Tsallis,
Introduction to Nonextensive Statistical Mechanics,
Springer, New York, 2009.


\bibitem{Bravo-PRD-2016}
R. Bravo, M. S. Plyushchay,
Phys. Rev. D 93 (2016) 105023.



\bibitem{RegoMonteiro-Nobre-2013-PRA}
M. A. Rego-Monteiro, F. D. Nobre,
Phys. Rev. A 88 (2013) 032105.


\bibitem{Hall-PRA-2000}
M. J. W. Hall,
Phys. Rev. A 62 (2000) 012107.


\bibitem{weizsacker-1935}
C. F. von Weizs\"acker,
Z. Phys. 96 (1935) 431.


\bibitem{Falaye-2016}
B. J. Falaye, F. A. Serrano, Shi-Hai Dong,
Phys. Lett. A 380 (2016) 267.


\bibitem{YanezNavarro-2014}
G. Ya\~nez-Navarro, Guo-Hua Sun, T. Dytrych,
K. D. Launey, Shi-Hai Dong, J. P. Draayer,
Ann. of Phys. 348 (2014) 153.


\bibitem{Macedo-Guedes-2015}
D. X. Macedo, I. Guedes,
Physica A 434 (2015) 211.


\bibitem{Furuichi-JMP}
S. Furuichi,
J. Math. Phys. 50 (2009) 013303.


\bibitem{Schmidt-2006}
A. G. M. Schmidt,
Phys. Lett. A 353 (2006) 459.


\bibitem{Lopez-Rosa-2011}
S. L\'opez-Rosa, ·J. Montero, P. S\'anchez-Moreno, J. Venegas, J. S. Dehesa,
J. Math. Chem. 49 (2011) 971.





\end{thebibliography}
\end{document}